\documentclass{article}
\usepackage{arxiv}

\usepackage{cmap}
\usepackage[T1]{fontenc}
\usepackage[utf8]{inputenc}
\usepackage{lmodern}
\usepackage{textcomp}

\usepackage{amsmath, amssymb, amsthm}
\usepackage{graphicx}
\usepackage{hyperref}
\usepackage[numbers]{natbib}
\usepackage{doi}
\usepackage[width=0.8\linewidth]{caption}
\usepackage{listings}
\usepackage{amsmath}
\usepackage{float}
\usepackage{xcolor}

\newtheorem{lemma}{Lemma}
\newtheorem{theorem}{Theorem}
\newtheorem{corollary}{Corollary}

\newtheorem{observation}{Observation}

\lstdefinestyle{python}{
  language=Python,
  basicstyle=\ttfamily,
  keywordstyle=\color{blue}\bfseries,
  stringstyle=\color{orange},
  commentstyle=\color{gray}\itshape,
  showstringspaces=false,
  frame=none,
  breaklines=true,
  tabsize=2
}

\title{Finding All Bounded-Length Simple Cycles in a Directed Graph --- Revisited}

\author{
  Frank Bauern\"oppel \\
  Computer Engineering\\
  Hochschule für Technik und Wirtschaft, Berlin, Germany\\
  \texttt{frank.bauernoeppel@htw-berlin.de}
  \And
  J\"org-R\"udiger Sack \\
  Carleton University Ottawa, School of Computer Science, Canada \\
  \texttt{sack@scs.carleton.ca}
}

\begin{document}
\maketitle

\begin{abstract}
In 2021, Gupta and Suzumura proposed a novel algorithm for enumerating all bounded-length simple cycles 
in directed graphs~\cite{DBLP:journals/corr/abs-2105-10094}. 
In this work, we present a concrete counter-example demonstrating that the proposed algorithm fails to enumerate certain valid cycles. 
Analyzing it, we pinpoint the precise step at which the original correctness proof breaks down.
We also identify a gap in the original proof of the delay bound claimed.
Finally, we propose algorithm \textsc{SimpleSearch} avoiding these flaws by construction, 
while achieving the delay bound $O(k(n + m))$ per cycle output or termination;
where $k$ is the length bound, $n$ the number of nodes, and $m$ the number of edges in the finite simple directed graph $G$.
\end{abstract}
 
\keywords{Graph Algorithms \and Cycle Enumeration \and Algorithm Analysis}

\section{Introduction}

\paragraph{Literature Review.}
The problem of enumerating all \emph{simple cycles} (cycles without repeated vertices) in a graph is fundamental in graph theory, with applications in areas such as chemistry, circuit analysis, and network science. 
Early work by Tiernan~\cite{tiernan1970} proposed a backtracking-based algorithm for finding elementary circuits, but it suffered from redundancy and exponential runtime in dense graphs. 
Johnson~\cite{DBLP:journals/siamcomp/Johnson75} later introduced the most influential algorithm for directed graphs, combining depth-first search with a blocking/unblocking mechanism 
to efficiently avoid revisiting fruitless paths. 

His algorithm runs in $O((c+1) \cdot (n+m))$ time, where $n$ is the number of vertices, $m$ the number of edges, and $c$ the number of simple cycles. 
Tarjan~\cite{tarjan1972} also contributed foundational work on strongly connected components, which remains an essential preprocessing step in many cycle-enumeration methods. 
Subsequent improvements, such as those by Szwarcfiter and Lauer~\cite{szwarcfiter1976}, refined Tiernan’s approach for better practical performance. 
Two more recent lines of work are closest in spirit to the present setting,
though each differs from it along an important dimension.

Birmelé et al.~\cite{birmele2013} give output-sensitive algorithms that list \emph{all}
simple cycles, and all $st$-paths, in \emph{undirected} graphs, optimal in the
total output size but with no bound on path or cycle length.
Rizzi et al.~\cite{DBLP:journals/corr/RizziSS14}, by contrast, do impose a length bound,
listing all length-bounded simple $st$-paths; their focus is the weighted case,
matching the classic $K$-shortest-paths time bounds in only $O(n+m)$ space, with
the unweighted case --- the one relevant here --- treated in a single remark that
we return to in \autoref{label:complexity}.
Grossi~\cite{DBLP:reference/algo/Grossi16} surveys enumeration algorithms in graphs.

The most recent algorithm for Finding All Bounded-Length Simple Cycles in a Directed Graph is due to 
Anshul Gupta and Toyotaro Suzumura~\cite{DBLP:journals/corr/abs-2105-10094}.
Their algorithm builds upon ideas of Johnson's algorithm.
But, instead of using a binary lock value for each node as in Johnson's algorithm, 
the new algorithm uses a numeric lock value for each node, which is designed to prevent futile searches when a cycle search gets too deep.

\paragraph{Our Contributions.}
In an experimental investigation of a GIS problem, we examined the Gupta-Suzumura algorithm. 
During initial testing, one of the expected cycles in a benchmark graph was not detected, leading us to suspect a possible flaw in our implementation. 
After carefully verifying the correctness of our code --- which was a line-by-line translation of the pseudocode provided in~\cite{DBLP:journals/corr/abs-2105-10094} 
into Python --- we proceeded to conduct a detailed analysis of the specific test case. 

This analysis revealed that the correctness proof in~\cite{DBLP:journals/corr/abs-2105-10094} rests on an invalid inference, and that the published pseudocode, 
when executed on the identified graph, indeed fails to enumerate one valid simple cycle in the test case, and often more in the general case. 

In this paper, we present a concrete counter-example to the algorithm of Gupta and Suzumura~\cite{DBLP:journals/corr/abs-2105-10094}. 
After introducing the terminology (\autoref{section:terminology}) and recalling the original algorithm (\autoref{section:original_algorithm}),
we provide a step-by-step examination of the algorithm's execution on this instance (\autoref{section:counter_example}), 
identify the precise step in the original correctness proof where the argument fails, and trace the omission to its root cause (\autoref{label:proof_analysis}): 
the algorithm's lock mechanism caches a backward distance that is valid only relative to one search stack and later reuses it under a different stack. 
In \autoref{label:revised}, we propose a corrected algorithm, \textsc{SimpleSearch}, that avoids this flaw by recomputing the relevant distance at every search level, and we establish its correctness. 
As we then show in \autoref{label:complexity}, the algorithm preserves the desired property of being sensitive to the number of simple cycles of a specified length.

\paragraph{On the choice of a redesign.}
Earlier versions of this work pursued a more conservative remedy: repairing the
original algorithm in place by modifying \textsc{RELAX\_LOCKS} to reset locks to infinity
rather than to a stack-relative value. Extensive testing indicates that this
modification does restore completeness, and we have found no counter-example to it.
We nonetheless do not adopt it, for two reasons.

First, we were unable to adapt the completeness proof for the in-place
fix. An argument following the structure of the original eventually relies on a
lock reset propagating back to the blocked node along a chain of \emph{Blist}
entries, a step whose validity depends on the search stack at the moment of
relaxation --- precisely the kind of stack-dependent reasoning whose failure we
diagnose in the original algorithm.

Second, retaining the lock architecture leaves the running time unresolved. The
delay analysis would inherit that of the original, which rests on the claim that
no vertex is revisited more than $k-1$ times before a cycle is found. That claim
is false: on a recently found instance (\autoref{appendix:visit_bound}), a vertex is
revisited well beyond this bound between two outputs, so the claimed per-output
delay does not follow as argued. This was a further reason not to attempt to
repair both the completeness and the delay proofs of the lock-based algorithm.

Rather than rest our result on an unproven completeness lemma and an unestablished
delay bound, we adopt \textsc{SimpleSearch}, which dispenses with lock maintenance
entirely and admits direct proofs of soundness, completeness, and per-output delay.

\section{Terminology}\label{section:terminology}

A \emph{directed graph} $G$ is defined as an ordered pair $(V, E)$, 
where $V=\{v_1, v_2, \dots, v_n\}$ denotes the set of \emph{nodes} (or vertices) 
and $E \subseteq V \times V$ denotes the set of \emph{directed edges}, represented as ordered pairs of nodes.  

A node $v_i \in V$ is said to be \emph{connected} to a node $v_j \in V$ by an edge $(v_i, v_j)$ 
if and only if the ordered pair $(v_i, v_j)$ belongs to $E$.  

A graph $G=(V,E)$ that has neither loops $(v,v) \in E$ nor multiple edges connecting the same ordered pair of nodes
is called a \emph{simple} graph.

A \emph{path} in $G$ is a finite sequence of nodes $(v_{i_1}, v_{i_2}, \dots, v_{i_m})$ such that $(v_{i_j}, v_{i_{j+1}}) \in E$ for all $1 \le j< m$. 

A \emph{simple path} is a path in which all nodes are distinct.  

A \emph{simple cycle} is a path  $(v_{i_1}, v_{i_2}, \dots, v_{i_m})$ with $m \ge 2$ such that  $(v_{i_1}, v_{i_2}, \dots, v_{i_{m-1}})$ is a simple path and $v_{i_1} = v_{i_m}$.  

The \emph{length} of a path or cycle $X$ is the number of edges it contains,
denoted $\|X\|$. We write $V(X)$ for the set of nodes occurring on $X$.

For brevity, edges, paths, and cycles are denoted by concatenating their node sequences.
In the graph $G$ from \autoref{figure:graph_g} for example, $AD$ is an edge, $ADB$ is a path, and $ADA$ is a cycle.

\section{The original Cycle Finding Algorithm from Gupta and Suzumura}\label{section:original_algorithm}

The original algorithm from  Gupta and Suzumura~\cite{DBLP:journals/corr/abs-2105-10094} consists of three functions: 
\begin{itemize}
  \item $\textsc{LC\_CYCLES}(G, k)$: The outer loop that initializes data structures and iteratively calls the cycle search function for each node $s$ in the graph $G$.
  \item $\textsc{CYCLE\_SEARCH}(G^s, s, k, flen)$: A modified depth-first search function that explores paths starting from a given node $s$ in a subgraph $G^s$ of $G$, looking for cycles of length up to $k$.
  \item $\textsc{RELAX\_LOCKS}(u, k, blen)$: A helper function that updates the lock values of nodes after a cycle is found, allowing them to be revisited in future searches.
\end{itemize}

For completeness, \autoref{appendix:pseudocode} presents the functions \textsc{CYCLE\_SEARCH} and \textsc{RELAX\_LOCKS} 
from~\cite{DBLP:journals/corr/abs-2105-10094}, translated into Python to demonstrate the counter-example.
Function \textsc{LC\_CYCLES} mainly manages an outer loop for finding more cycles not containing the first start node $s$, and is of no further relevance here.

\section{A Counter-Example to \textsc{CYCLE\_SEARCH} Completeness}\label{section:counter_example}

\begin{figure}[]
    \centering
    \includegraphics[width=0.75\textwidth]{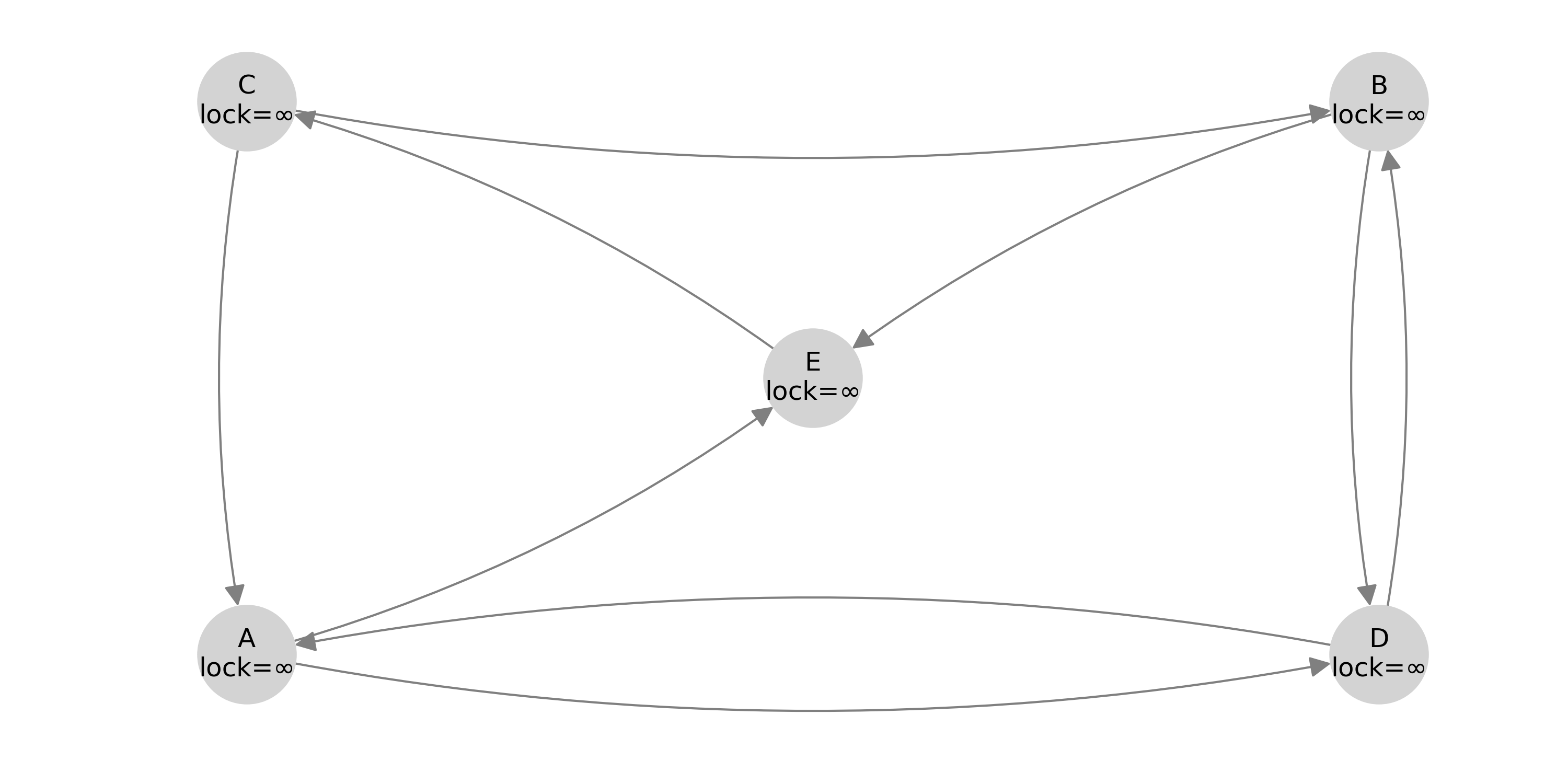}
    \caption{Directed graph $G = (V, E)$}\label{figure:graph_g}
\end{figure}

Consider the directed graph $G = (V, E)$ with 
\begin{itemize}
  \item node set $V = \{A, B, C, D, E\}$ and
  \item edge set $E = \{AD, AE, BD, BE, CA, CB, DA, DB, EC\}$
\end{itemize}
The graph $G$ is depicted in \autoref{figure:graph_g}.

In an internal computer representation of the graph using an adjacency list, the edges are stored in the order shown above. 
This ordering affects the sequence in which nodes are explored during depth-first searches. 
In theory, the ordering should not affect the set of cycles found, but here it does, as will be clear later.

It is straightforward to see that graph $G$ contains the following simple cycles starting at the first node $A$ (in lexicographical order):
\begin{itemize}
  \item $ADA$ (length 2)
  \item $ADBECA$ (length 5)
  \item $AECA$ (length 3)
  \item $AECBDA$ (length 5)
\end{itemize}

The two longest cycles are Hamiltonian and therefore $G$ is strongly connected.

Let the function \textsc{LC\_CYCLES}(G, k) from~\cite{DBLP:journals/corr/abs-2105-10094}, the outer loop, 
be called with the above graph $G$ and $k=5$ as the length bound parameter.

Function \textsc{LC\_CYCLES} starts with initializing for each node $v$ in $G$
\begin{itemize}
  \item \emph{$lock[v]$} to infinity and
  \item \emph{$Blist[v]$} to an empty set. 
\end{itemize}
The $lock$s are numeric values designed to block nodes from being revisited during the search under certain conditions.
The $Blist$s are of no importance here, as they will remain empty throughout the algorithm execution in this example, 
but are needed in the general case.

After initialization, the first call to \textsc{CYCLE\_SEARCH}(Gs, s, k, 0) is made in line 9 of \textsc{LC\_CYCLES} 
with the parameters $s=A$ (the first node of graph $G$), $Gs = G$, and $k=5$.

In the following sub-sections, a commented trace  of the 
execution of \textsc{CYCLE\_SEARCH}(G, A, 5, 0) is given.

\subsection{Finding Cycle $ADA$}

The search starts as follows:

\begin{figure}[H]
\begin{verbatim}
  1: cycle_search stack=      v='A' k=5 flen=0: push A, blen←inf, lock[A]←0
  2: cycle_search stack=A     v='D' k=5 flen=1: push D, blen←inf, lock[D]←1
  3: cycle_search stack=AD    v='D' ##### cycle ADA   found, blen←1 #####
\end{verbatim}
\end{figure}

\begin{figure}[H]
    \centering
    \includegraphics[width=0.75\textwidth]{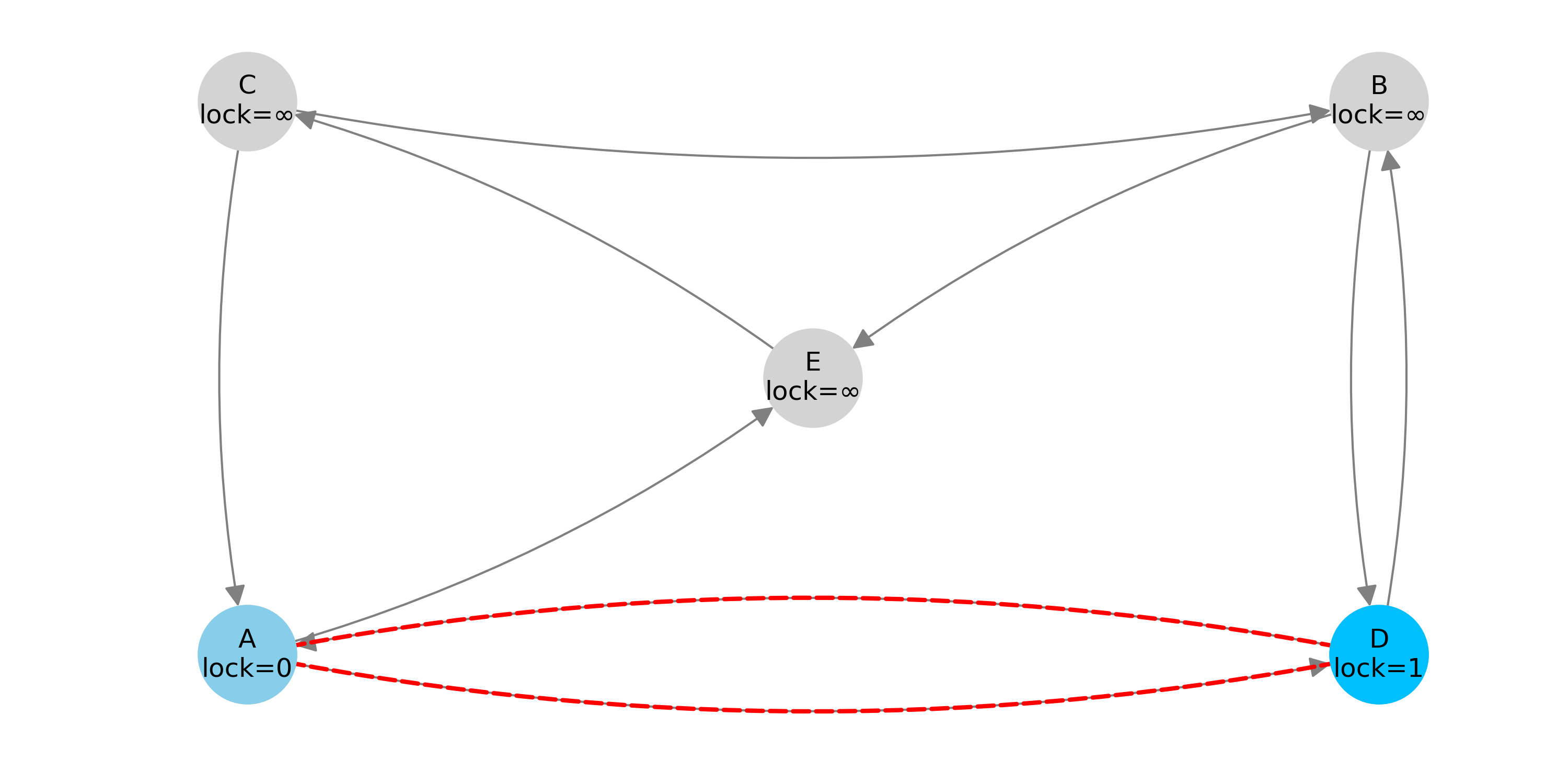}
    \caption{Graph $G$ after first DFS in \textsc{CYCLE\_SEARCH}(G, A, 5, 0) found cycle $ADA$.}\label{figure:graph_g_1}
\end{figure}

The result is depicted in \autoref{figure:graph_g_1}.

\subsection{Finding Cycle $ADBECA$}

\begin{figure}[H]
    \centering
    \includegraphics[width=0.75\textwidth]{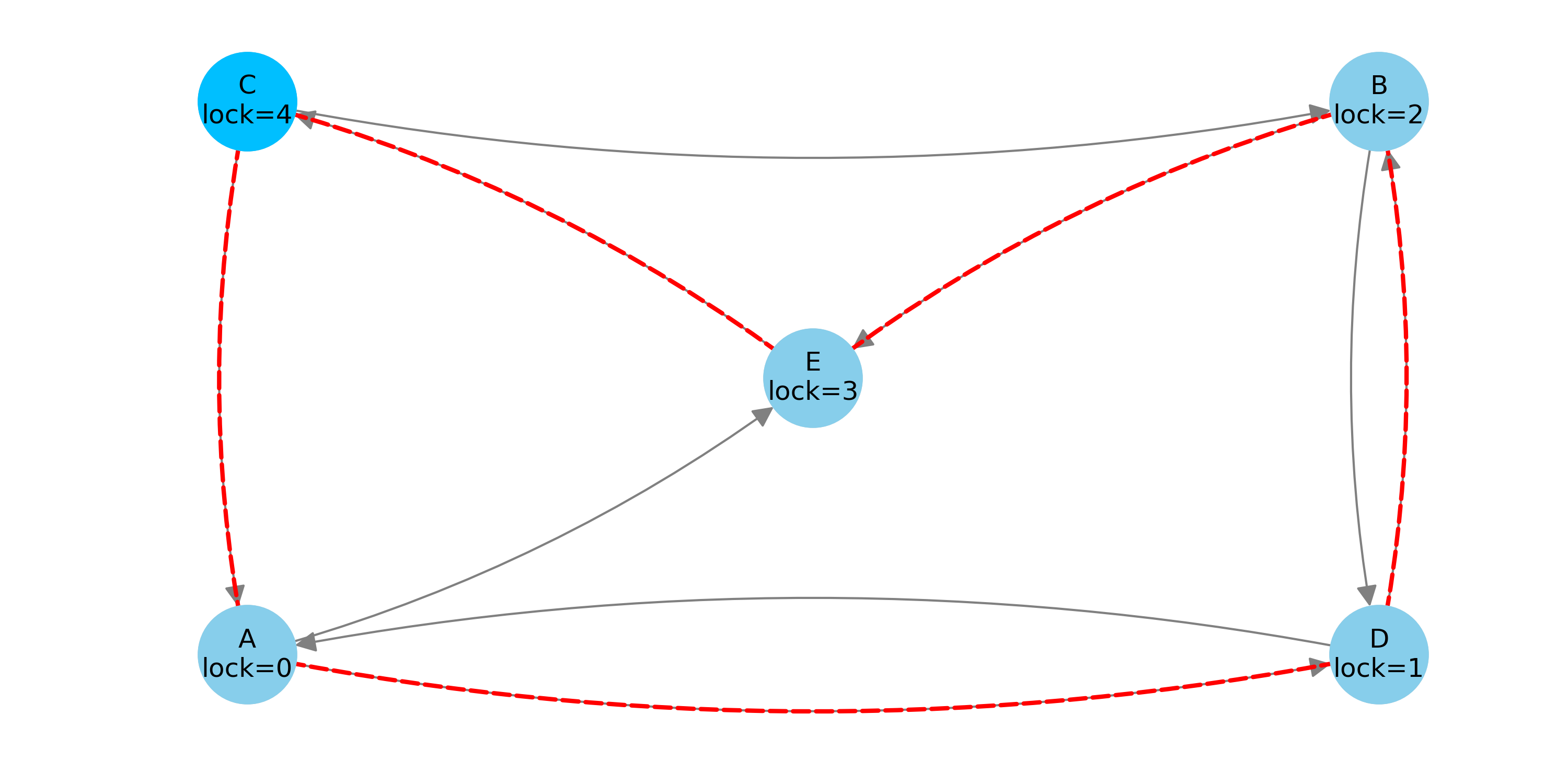}
    \caption{Graph $G$ after second DFS in \textsc{CYCLE\_SEARCH}(G, A, 5, 0) found cycle $ADBECA$.}\label{figure:graph_g_2}
\end{figure}

After backtracking to $D$, the next cycle $ADBECA$ is found when the  depth-first-search continues:

\begin{figure}[H]
\begin{verbatim}
  4: cycle_search stack=AD    v='B' k=5 flen=2: push B, blen←inf, lock[B]←2
  5: cycle_search stack=ADB   v='E' k=5 flen=3: push E, blen←inf, lock[E]←3
  6: cycle_search stack=ADBE  v='C' k=5 flen=4: push C, blen←inf, lock[C]←4
  7: cycle_search stack=ADBEC v='C' ##### cycle ADBECA found, blen←1 #####
\end{verbatim}
\end{figure}

see \autoref{figure:graph_g_2}.

During the search from $D$, the locks of the visited nodes $B$, $E$, and $C$ 
are set to their index (depth) in the search stack. 
All other locks remain unchanged.

Backtracking finally returns to $A$, thereby calling 
\textsc{RELAX\_LOCKS}(u, k, blen) for each node visited, in backtracking order.

\begin{figure}[H]
\begin{verbatim}
  8:  relax_locks stack=ADBEC v='C' k=5 blen=1 lock[C]=4
  9:  relax_locks stack=ADBEC v='C' k=5 blen=1 lock[C]←5 (=k-blen+1)
 10: cycle_search stack=ADBEC v='C' flen=4 lock[C]=5: pop C, return blen=1
 11: cycle_search stack=ADBE  v='E' blen←2
 12:  relax_locks stack=ADBE  v='E' k=5 blen=2 lock[E]=3
 13:  relax_locks stack=ADBE  v='E' k=5 blen=2 lock[E]←4 (=k-blen+1)
 14: cycle_search stack=ADBE  v='E' flen=3 lock[E]=4: pop E, return blen=2
 15: cycle_search stack=ADB   v='B' blen←3
 16:  relax_locks stack=ADB   v='B' k=5 blen=3 lock[B]=2
 17:  relax_locks stack=ADB   v='B' k=5 blen=3 lock[B]←3 (=k-blen+1)
 18: cycle_search stack=ADB   v='B' flen=2 lock[B]=3: pop B, return blen=3
 19: cycle_search stack=AD    v='D' blen←1
 20:  relax_locks stack=AD    v='D' k=5 blen=1 lock[D]=1
 21:  relax_locks stack=AD    v='D' k=5 blen=1 lock[D]←5 (=k-blen+1)
 22: cycle_search stack=AD    v='D' flen=1 lock[D]=5: pop D, return blen=1
 23: cycle_search stack=A     v='A' blen←2
\end{verbatim}
\end{figure}

Note step 17 in particular: $B$ receives $blen=3$ and hence $lock[B]=3$, 
recording a backward distance of 3 even though a length-2 return $BDA$ exists in the graph.
This is the value that will later block $AECBDA$.

\begin{figure}[H]
    \centering
    \includegraphics[width=0.75\textwidth]{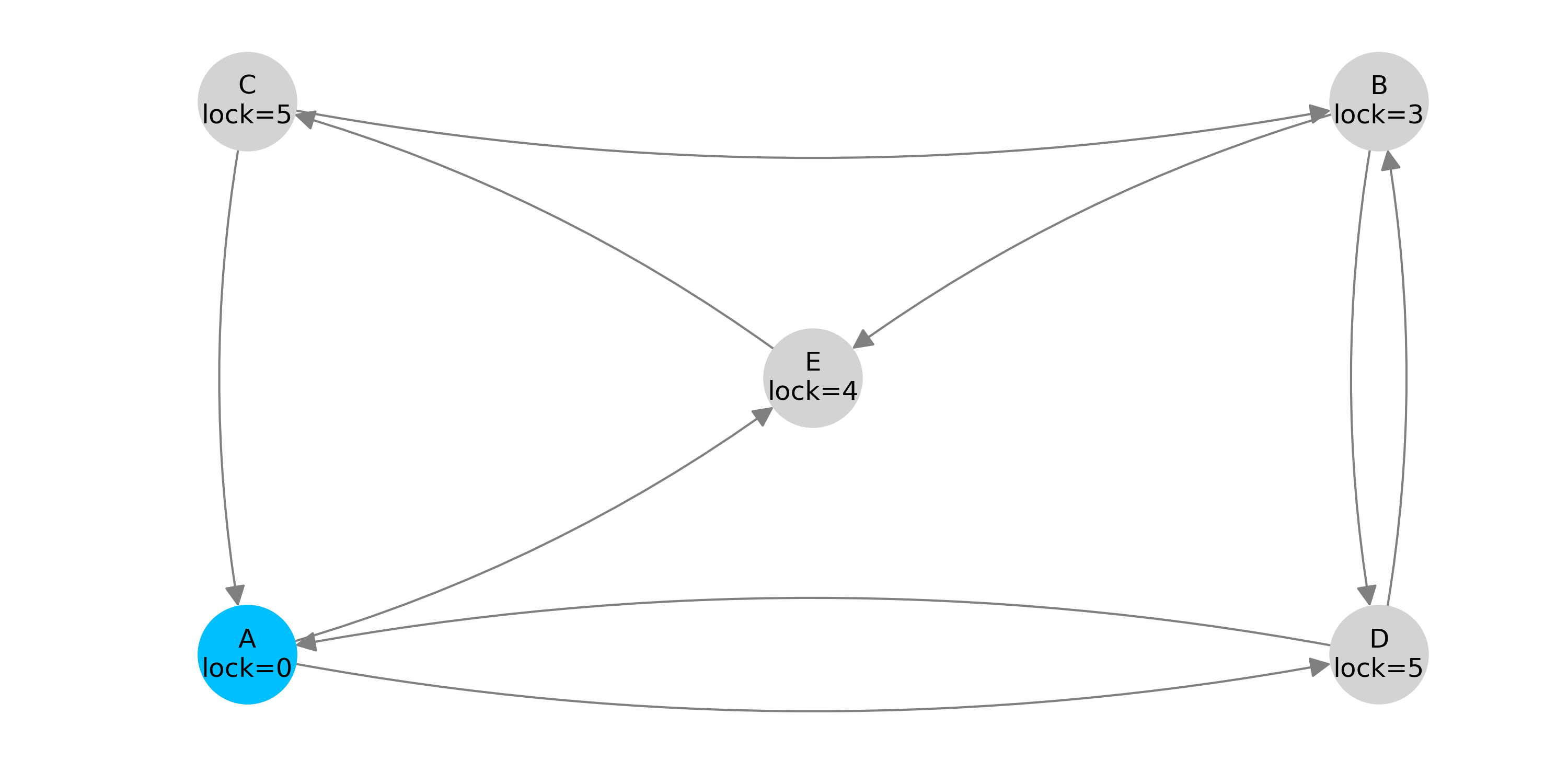}
    \caption{Graph $G$ after backtracking to $A$ in \textsc{CYCLE\_SEARCH}(G, A, 5, 0).}\label{figure:graph_g_3}
\end{figure}

In these calls, the $blen$ parameter is representing the backward distance to $A$ along the cycle found. 
Since all $Blist$ are still empty sets, \textsc{RELAX\_LOCKS} never recurses to itself.
\textsc{RELAX\_LOCKS} will possibly increase a lock value of a node when $blen$ is sufficiently small. 
After backtracking to $A$, graph $G$ will look as shown in \autoref{figure:graph_g_3}.

\subsection{Finding Cycle $AECA$}

Now, the depth-first search finds cycle $AECA$. 
The locks of nodes $E$ and $C$ are set to their index (depth) in the third search stack. 

\begin{figure}[H]
\begin{verbatim}
 24: cycle_search stack=A     v='E' k=5 flen=1: push E, blen←inf, lock[E]←1
 25: cycle_search stack=AE    v='C' k=5 flen=2: push C, blen←inf, lock[C]←2
 26: cycle_search stack=AEC   v='C' ##### cycle AECA  found, blen←1 #####
\end{verbatim}
\end{figure}

\begin{figure}[H]
    \centering
    \includegraphics[width=0.75\textwidth]{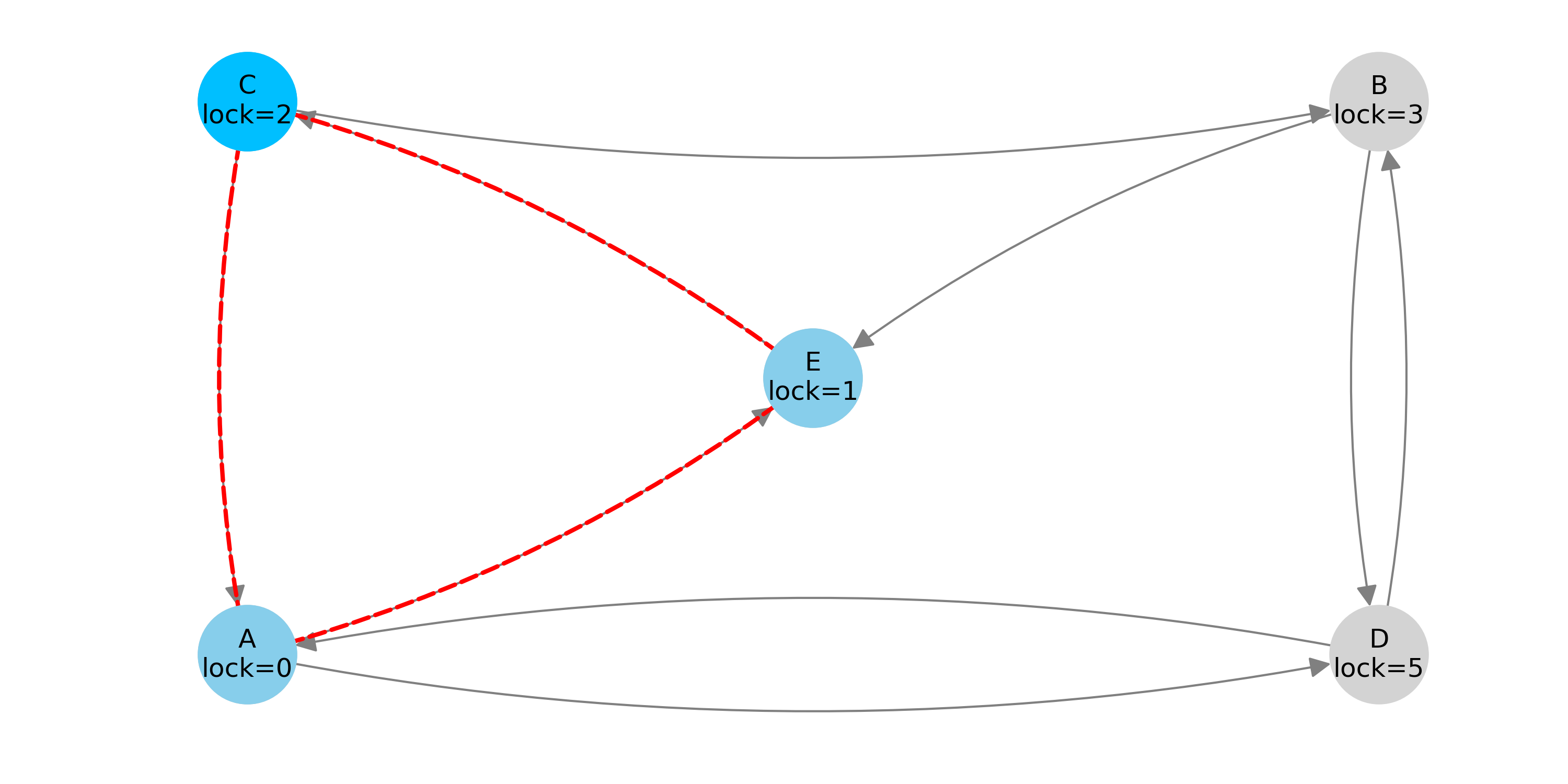}
    \caption{
      Graph $G$ after third DFS in \textsc{CYCLE\_SEARCH}(G, A, 5, 0) found cycle $AECA$.
      Node $B$ (grey, not on the stack) retains $lock=3$
      from the earlier visit and will later block the edge $C\to B$.
    }\label{figure:graph_g_4}
\end{figure}

The other nodes are not affected, see \autoref{figure:graph_g_4}.

\subsection{Missing Cycle $AECBDA$}

After cycle $AECA$ was found, but before node $C$ is completely explored, node $B$ is considered in the search. 
But, as the lock of node $B$ has a value of 3 (see steps 17, 18), according to line 9 of the pseudocode, 
the search does \emph{not} recurse to $B$. 

\begin{figure}[H]
\begin{verbatim}
 27:  relax_locks stack=AEC   v='C' k=5 blen=1 lock[C]=2
 28:  relax_locks stack=AEC   v='C' k=5 blen=1 lock[C]←5 (=k-blen+1)
 29: cycle_search stack=AEC   v='C' flen=2 lock[C]=5: pop C, return blen=1
 30: cycle_search stack=AE    v='E' blen←2
 31:  relax_locks stack=AE    v='E' k=5 blen=2 lock[E]=1
 32:  relax_locks stack=AE    v='E' k=5 blen=2 lock[E]←4 (=k-blen+1)
 33: cycle_search stack=AE    v='E' flen=1 lock[E]=4: pop E, return blen=2
 34: cycle_search stack=A     v='A' blen←2
 35:  relax_locks stack=A     v='A' k=5 blen=2 lock[A]=0
 36:  relax_locks stack=A     v='A' k=5 blen=2 lock[A]←4 (=k-blen+1)
 37: cycle_search stack=A     v='A' flen=0 lock[A]=4: pop A, return blen=2
 38: halt
\end{verbatim}
\end{figure}

As a consequence, the proposed search algorithm misses cycle $AECBDA$ --- 
the recorded distance of 3 is too pessimistic, a point we analyze in \autoref{label:proof_analysis}.

\begin{figure}[H]
    \centering
    \includegraphics[width=0.75\textwidth]{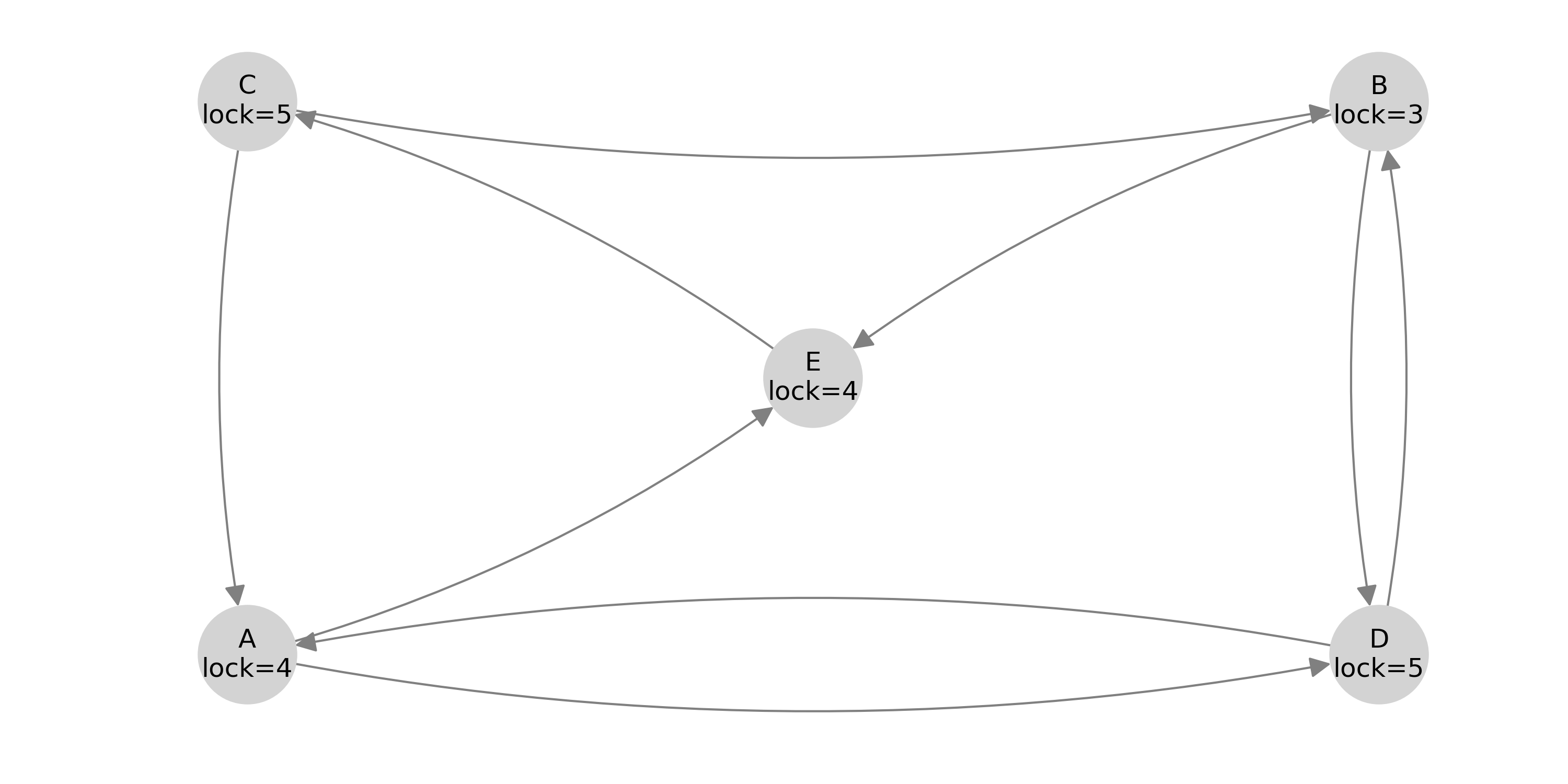}
    \caption{Graph $G$ after backtracking to $A$ in \textsc{CYCLE\_SEARCH}(G, A, 5, 0).}\label{figure:graph_g_5}
\end{figure}

After missing that cycle, the search backtracks from $C$ via $E$ to $A$, thereby updating (relaxing) the locks along the way.
The final result is shown in \autoref{figure:graph_g_5}.

Finally, \textsc{CYCLE\_SEARCH}(G, A, 5, 0) completes and returns to \textsc{LC\_CYCLES}.
In \textsc{LC\_CYCLES}, the outer loop, the first start node $s=A$ will now be removed from the graph and 
in all further calls to \textsc{CYCLE\_SEARCH}, node $A$ is no longer contained. 
Thus, cycle $AECBDA$ cannot be found anymore.

This example  demonstrates that the proposed algorithm does not always find all cycles of specified length $k$ in a directed graph.

\subsection{Remarks}

This counter-example was among the first identified through a systematic search of small, strongly connected directed graphs. 
An arbitrary number of additional counter-examples can be generated by introducing further nodes and edges 
connected to $G$ exclusively via edges outgoing from $D$ and edges incoming to $A$, adding more instances of the above construct, etc.

Although modifying the order of edges in the adjacency list --- that is, the internal representation of the graph --- 
may enable the algorithm to enumerate all cycles, this does not constitute a valid solution. 
A correct algorithm must identify all cycles independently of the order in which edges are presented in the input graph. 
Trying all orders is obviously infeasible.

\section{Analysis of the Completeness Proof}\label{label:proof_analysis}
The interesting part in the correctness proof in~\cite{DBLP:journals/corr/abs-2105-10094}, chapter 4, is 
Lemma 2: Function \textsc{CYCLE\_SEARCH} outputs every cycle of length $k$ or less in subgraph $G^s$.
We first locate the precise inferential step at which the proof breaks down, then
identify the underlying design flaw that causes it.

\paragraph{The precise failure.}
The first valid (length-$k$-bounded) cycle not detected by the algorithm is $AECBDA$.
In the terminology of the original paper, edge $CB$ is the first \emph{blocked} edge,
so $v_{i+1}=B$ with $i=2$ and the missed cycle has length $l+1=5$, i.e.\ $l=4$.
Search stack $AEC$ is not the first visit to node $B$; the last visit before it
occurred with search stack $AD$, during the second depth-first search, which
yielded cycle $ADBECA$. Since that last visit did find a cycle, the \emph{second
possibility} in the proof of Lemma 2 applies, and \textsc{RELAX\_LOCKS} was called
on $B$.

The proof of this possibility (pages 7--8 of~\cite{DBLP:journals/corr/abs-2105-10094})
argues that, because the missed cycle contains a path of length $l-i$ from
$v_{i+1}$ to $s$, this path ``must have been successfully traversed during the
last visit to $v_{i+1}$'', and concludes inequality~(5), $b \le l-i$, where $b$
is the backward distance recorded at $v_{i+1}$ on that visit. This inference is
invalid. The \emph{existence} of the suffix $v_{i+1}\to\cdots\to v_l\to s$ is a
property of the graph, but whether it was \emph{traversable} on the last visit
depends on the search stack at that visit: any interior vertex of the suffix that
was on the stack is blocked, rendering the path unavailable. The backward
distance $b$ actually recorded is therefore not the length of the graph-shortest
suffix, but the length of the shortest suffix \emph{whose interior avoids the
then-current stack}, which may be strictly larger; so~(5) can fail.

In our instance the suffix the proof relies on is $B\to D\to A$ of length
$l-i=2$, but its interior vertex $D$ was on the stack $AD$ and hence blocked, so
$BDA$ was not traversable. The visit instead returned along $B\to E\to C\to A$ of
length $3$, recording $b=3$ (steps 15--18 of the trace). Thus $b=3 > l-i=2$,
inequality~(5) is false, the lock of $B$ is set to $k-b+1=3$, and the later
arrival at $B$ with $\mathit{flen}=i+1=3$ is blocked because the admission test
$\mathit{flen}+1<\mathit{Lock}(B)$, i.e.\ $3<3$, is false. The cycle $AECBDA$ is
lost. This establishes that Lemma~2
of~\cite{DBLP:journals/corr/abs-2105-10094} contains a gap.

\paragraph{Root cause.}
The error is not an isolated arithmetic slip but a conflation of two distinct
notions of distance. The lock value $k-b+1$ caches a \emph{stack-relative}
quantity --- the backward distance achievable from a vertex under one particular
search stack --- and the algorithm later consults it under a \emph{different}
stack, as though it were a graph-relative invariant. When the genuine shortest
suffix of a future cycle happened to be blocked at the recording visit, the
cached lock is too restrictive, and a valid cycle is lost. The \textsc{SimpleSearch}
algorithm of \autoref{label:revised} avoids this structurally: the quantity
governing admission, a relative distance to $s$, is recomputed
against the current search path at every level and is never carried across
stacks.

\section{The \textsc{SimpleSearch} Algorithm}\label{label:revised}

In \autoref{figure:simplesearch} we present \textsc{SimpleSearch} for the bounded-length simple cycle enumeration
problem: enumerate all simple cycles through a fixed node $s$ of length at most
$k$. A cycle is reported as the open node sequence $(s,v_1,\dots,v_h)$ of its
search path, with the closing node $s$ left implicit; we identify such an output
with the simple cycle $(s,v_1,\dots,v_h,s)$ it denotes.

In contrast to the barrier-caching algorithm of~\cite{DBLP:journals/corr/abs-2105-10094},
\textsc{SimpleSearch} performs no barrier bookkeeping at all. Instead, before
each recursion it recomputes, by a single depth-limited breadth-first search in
the reverse graph from the start node $s$, exactly which successors of the
current node can still reach $s$ within the remaining budget. This trades the
(delicate, and as we have seen incorrect) lock maintenance for one linear
traversal per search level. As we show below, the resulting algorithm is sound,
complete, and has per-output delay $O(k\,(n+m))$.

Neither the delay bound we obtain nor the $O(n+m)$ space complexity is entirely new.
For listing $k$-bounded simple $st$-paths in unweighted directed graphs,
Rizzi et al.~\cite{DBLP:journals/corr/RizziSS14} already achieve $O(k\,m)$ delay in
$O(n+m)$ space: their binary-partition scheme guards each recursive call by a
single-source shortest-path test --- a breadth-first search in $O(m)$ time in the
unweighted case --- over a recursion tree of height bounded by $k$, and it stays
within linear space because each call stores only its difference from the parent's
graph rather than a copy. The bounded-length $s$-cycle problem reduces to
bounded-length $st$-path listing by splitting $s$ into an out-copy and an in-copy
and listing the paths between them, so \textsc{SimpleSearch} matches this route in
both delay and space.

We do not take that route, and the difference is one of method rather than of
resources. Their partition removes a vertex and recurses on the reduced graph;
\textsc{SimpleSearch} leaves the input graph untouched and mutates only the search
path, filtering the successors of the current node against a path-relative backward
reach recomputed at each level. Working directly on cycles rather than through the
split-vertex reduction lets us state the algorithm and its proofs over a single
fixed graph, with no auxiliary construction and no cycle-to-path bijection to
maintain; the soundness, completeness, and delay arguments below are correspondingly
self-contained and specific to the cycle problem. The match
with~\cite{DBLP:journals/corr/RizziSS14} is thus one of asymptotic regime --- delay
and space alike --- reached by a direct, cycle-specific formulation that also serves
as a corrected counterpart to \textsc{CYCLE\_SEARCH}.

\begin{figure}[H]
\small
\begin{lstlisting}[style=Python]
from collections import deque

def simple_search(G, s, k):
    """Enumerate all simple cycles in G of bounded length k containing node s."""

    def reach(blocked, successors, budget):
        queue = deque()
        queue.append((s, 0))
        reached = {s}
        while queue:
            (u, d) = queue.popleft()
            if d >= budget:
                break
            for v in G.predecessors(u):
                if v not in blocked and v not in reached:
                    reached.add(v)
                    queue.append((v, d + 1))
        return [w for w in successors if w in reached]

    def search(path, v, budget):
        if path and v==s:
            print(path) # output cycle
        else:
            path.append(v)
            fruitful = reach(path, G.successors(v), budget - 1)
            for w in fruitful:
                search(path, w, budget - 1)
            path.pop() # v

    path = list()
    search(path, s, k)
\end{lstlisting}
\caption{The \textsc{SimpleSearch} algorithm.}\label{figure:simplesearch}
\end{figure}

In words, the algorithm grows a simple path from $s$ and yields it whenever it
closes back to $s$. The search starts with the path consisting of $s$ alone and
the full budget $k$; the path itself records which nodes are in use, so that none
is reused as an interior node. At each node reached, before descending any further,
the algorithm determines which nodes can reach $s$ within the remaining budget: 
it runs a breadth-first search \emph{backward} from $s$ --- that is, following edges in
reverse --- never passing through a node already on the path and never going
deeper than the remaining budget allows. Among the successors of the current node it then
keeps only those that this backward search found, the \emph{fruitful} ones; a
successor that cannot reach $s$ within the remaining budget is discarded at once,
so no step is taken that could not be completed to a valid cycle. Each fruitful
successor is handed to a recursive call. If that successor is $s$, the recursion
yields the current path --- the cycle just closed --- and returns without extending
it; otherwise the successor is appended, the search recurses with the budget reduced
by one, and on return the successor is removed again, freeing it for other branches.
Because the backward search is exact rather than approximate, every recursive call
leads to at least one yielded cycle: the algorithm never descends into a branch that
yields nothing. This is the property that underlies the
per-output delay bound, made precise in \autoref{label:complexity}.

\paragraph{Implementation Notes}
The listing uses Python sets and a list membership test for readability. A
\texttt{set} offers only expected constant-time membership under hashing, which is
not a worst-case guarantee; likewise testing $w\in\texttt{path}$ on a list is
linear. An implementation therefore backs both the reachable set and the path by
boolean arrays indexed by node, so that every membership test is worst-case
constant time. We assume this representation throughout 
while omitting it in \autoref{figure:simplesearch} for clarity.

\subsection{Preliminaries}

We reuse the notation of \autoref{section:terminology}. The search maintains a
\emph{search path} \texttt{path}, a list of nodes; while \texttt{search} is
active on it, this list realizes a simple path
$P=(s=v_0,v_1,\dots,v_h)$ in $G$ starting at $s$, of length $\|P\|=h$ and
with node set $V(P)$.
The list \texttt{path}
serves both as the growing path and as the set of nodes to avoid: a node is
treated as forbidden exactly when it already lies on \texttt{path}. As noted
above, an output
$(s,v_1,\dots,v_h)$ is identified with the simple cycle
$(s,v_1,\dots,v_h,s)$ it closes.

The execution of \texttt{search} forms a rooted \emph{call tree} $T$: its nodes are
the \texttt{search} calls, its root is the initial call (on the empty path, with
budget $k$), and the children of a call are the recursive calls it makes, in
execution order. We use the usual tree vocabulary --- \emph{parent}, \emph{child},
\emph{subtree}, \emph{leaf}, and \emph{depth} (the root has depth~$0$). Each call is
of one of two kinds, according to the branch it takes: a \emph{branching call} (the
\texttt{else} branch) appends its node, runs one \texttt{reach}, and recurses on the
fruitful successors, while an \emph{output call} (the \texttt{if} branch, taken when
\texttt{path} is non-empty and $v=s$) yields the current path and returns without
recursing. An output call is thus a leaf and runs no \texttt{reach}; the output
calls correspond exactly to the outputs.

For a node set $S\subseteq V$ and the fixed target $s$ we use the \emph{relative
distance}
\[
  \mathrm{dist}_S(x,s)=\min\{\,\|Q\| : Q \text{ is an } x\text{-}s
  \text{ path with } V(Q)\cap S\subseteq\{x,s\}\,\},
\]
with $\mathrm{dist}_S(x,s)=\infty$ if no such path exists: the length of a
shortest $x$-to-$s$ path whose interior avoids $S$. In particular
$\mathrm{dist}_S(s,s)=0$.

\begin{observation}[Backward Reach]\label{obs:reach}
  Let $G$ be a simple directed graph with $n$ nodes and $m$ edges, 
  $s$ a node of $G$, 
  and $0<k\le n$ a given length bound.
  In a call \texttt{reach(path,\,successors,\,budget)} 
  (the search path is passed as the \texttt{blocked} argument) made while the search path
  is $P$, the \texttt{reached} set $R$ computed internally satisfies
  \[
    R\;=\;\{s\}\;\cup\;\{\,w\in V\setminus V(P) : \mathrm{dist}_{V(P)}(w,s)\le \texttt{budget}\,\},
  \]
  and the call returns the members of \texttt{successors} that lie in $R$, in their original order.
\end{observation}
\begin{proof}
  The traversal is a breadth-first search in the reverse graph, started at $s$.
  Its guard \texttt{v not in blocked} prevents any node of $V(P)$ from being
  enqueued, so no node of $V(P)$ other than $s$ is ever recorded; and it
  expands no node once its depth reaches \texttt{budget} (the guard
  \texttt{d\,>=\,budget} breaks before expansion). For a node $q\notin V(P)$, $q$
  is recorded with label $d$ iff $d$ is the length of a shortest $s$-to-$q$ path
  in the reverse graph all of whose vertices, apart from the endpoints, lie
  outside $V(P)$ --- equivalently a shortest $q$-to-$s$ path in $G$ whose interior
  avoids $V(P)$, i.e.\ $d=\mathrm{dist}_{V(P)}(q,s)$ --- and only such nodes with
  $d\le\texttt{budget}$ are recorded.
  Hence the computed set \texttt{reached} equals $R$, and the returned list
  consists of exactly those \texttt{successors} belonging to $R$, in list order.
\end{proof}

\subsection{Soundness}

We show that the algorithm is total and that every output is a valid length-$k$ bounded simple cycle.

\begin{observation}[Simple Search Path]\label{obs:simple-path}
  Let $G$ be a simple directed graph with $n$ nodes and $m$ edges,
  $s$ a node of $G$, and $0<k\le n$ a given length bound. The branching and output
  calls defined in the Preliminaries satisfy the following.
  \begin{itemize}
    \item In a branching call, after appending $v$ the search path
      $P=(s=v_0,\dots,v_h=v)$ is a simple path in $G$ with
      $\texttt{budget}=k-\|P\|\ge 1$, and distinct branching calls have distinct
      paths. Every branching call other than the initial one has $v_h\neq s$.
    \item In an output call, writing $P=(s=v_0,\dots,v_h)$ for \texttt{path}, we
      have $(v_h,s)\in E$ and $\texttt{budget}=k-\|P\|-1\ge 0$, so $\|P\|\le k-1$.
  \end{itemize}
\end{observation}
\begin{proof}
  By induction on the recursion. The initial call has empty \texttt{path} and so is
  a branching call; appending $s$ gives the path $(s)$ with $\|P\|=0$ and
  $\texttt{budget}=k\ge 1$. A branching call with path $P$ and
  $\texttt{budget}=k-\|P\|$ recurses, for each $w\in\texttt{fruitful}$, into
  \texttt{search(path,\,$w$,\,budget$-1$)}.

  If $w=s$ --- possible, as $s\in\texttt{reached}$ always --- then \texttt{path} is
  non-empty and the child is an output call: it leaves \texttt{path} unchanged, its
  last node $v_h$ has $(v_h,s)\in E$ since $s$ is a successor of $v_h$, and its
  budget is $k-\|P\|-1\ge 0$ because $\texttt{budget}=k-\|P\|\ge 1$.

  If $w\neq s$, then by \autoref{obs:reach} $w\in\texttt{reached}$ forces
  $\mathrm{dist}_{V(P)}(w,s)\le\texttt{budget}-1$; as $w\neq s$ this requires
  $\texttt{budget}-1\ge 1$, and $w\notin V(P)$. The child is a branching call: after
  it appends $w$ the path is again simple from $s$, with edge count $\|P\|+1$ and
  budget $\texttt{budget}-1=k-(\|P\|+1)$, and current node $w\neq s$. 
  Each branching call extends its parent's path by a unique successor absent from it, 
  so branching-call paths are pairwise distinct and remain simple.
\end{proof}

\begin{theorem}[Soundness]\label{thm:ss-sound}
  Every sequence output by $\textsc{simple\_search}(G,s,k)$ denotes a
  simple cycle through $s$ in $G$ of length at most $k$.
\end{theorem}
\begin{proof}
  By \autoref{obs:simple-path} an output is produced by an output call yielding its
  path $P=(s=v_0,\dots,v_h)$, where $(v_h,s)\in E$ and $\|P\|=h\le k-1$. Hence
  $(s,v_1,\dots,v_h,s)$ is a simple cycle through $s$ of length $h+1\le k$.
\end{proof}

\begin{observation}[Termination]\label{obs:ss-terminate}
  $\textsc{simple\_search}(G,s,k)$ terminates.
\end{observation}
\begin{proof}
  Each \texttt{reach} visits every node and edge at most once, hence terminates. By
  \autoref{obs:simple-path} distinct branching calls have distinct simple paths, and
  a finite graph has finitely many simple paths, so there are finitely many
  branching calls; each spawns finitely many children (output calls are leaves), so
  the recursion is finite.
\end{proof}

\subsection{Completeness}

We show that every valid length-$k$ bounded simple cycle is produced as output exactly once.

\begin{lemma}[Call Completeness]\label{lem:ss-complete-call}
  Consider a branching call with search path $P=(s=v_0,\dots,v_h)$ and
  $\texttt{budget}=k-h$. Then \textsc{simple\_search} yields every simple cycle
  through $s$ of length at most $k$ that begins with the prefix $(s,v_1,\dots,v_h)$.
\end{lemma}
\begin{proof}
  By induction on $k-\|P\|=k-h\ge 1$. Let
  $C=(s=v_0,\dots,v_h,v_{h+1},\dots,v_{\ell-1},v_\ell=s)$ be a simple cycle through
  $s$ of length $\ell\le k$ that begins with the prefix $(s,\dots,v_h)$, and put
  $w=v_{h+1}$ (if $h=\ell-1$, then $w=v_\ell=s$).

  \emph{Case $w=s$.} Then $\ell=h+1$ and $(v_h,s)\in E$, so $s$ is a successor of
  $v_h$; since $s\in\texttt{reached}$ always, $s\in\texttt{fruitful}$, and the call
  recurses into \texttt{search(path,\,$s$,\,budget$-1$)}, an output call that yields
  $(s,\dots,v_h)=C$.

  \emph{Case $w\neq s$.} Since $C$ is simple, $w\notin V(P)$ and $(v_h,w)\in E$. The
  suffix $(w,\dots,s)$ is a $w$-$s$ path of length $\ell-h-1$ whose interior, by
  simplicity of $C$, avoids $\{v_0,\dots,v_h\}=V(P)$; hence
  $\mathrm{dist}_{V(P)}(w,s)\le\ell-h-1$. The call runs \texttt{reach} with
  $\texttt{budget}-1=k-h-1\ge\ell-h-1\ge\mathrm{dist}_{V(P)}(w,s)$, so by
  \autoref{obs:reach} (with $w\notin V(P)$) $w\in\texttt{reached}$. The call
  therefore recurses into the branching call with path $(s,\dots,v_h,w)$ and budget
  $k-(h+1)$. Since $C$ begins with this prefix and $k-(h+1)<k-h$, the induction
  hypothesis yields $C$.
\end{proof}

\begin{theorem}[Completeness]\label{thm:ss-complete}
  For a simple directed graph $G$ with $n$ nodes and $m$ edges and length bound $0<k\le n$,
  $\textsc{simple\_search}(G,s,k)$ outputs every simple cycle through $s$ of length at most $k$ in $G$.
\end{theorem}
\begin{proof}
  The initial call is a branching call with path $(s)$, with which every cycle
  through $s$ begins; apply \autoref{lem:ss-complete-call}.
\end{proof}

By \autoref{obs:simple-path} distinct branching calls have distinct paths, and each
output call yields the path of its parent branching call; since a branching call
recurses on the successor $s$ at most once, no cycle is yielded twice.

\subsection{Delay Bound}\label{label:complexity}

We work in the call tree $T$ of the Preliminaries. By \autoref{obs:simple-path} a
branching call has $\texttt{budget}=k-\|P\|\ge 1$, so $\|P\|\le k-1$, and its depth
in $T$ equals $\|P\|$; hence branching calls have depth at most $k-1$, and any
root-to-node descending path of $T$ contains at most $k$ branching calls, thus at
most $k$ \texttt{reach} runs.

The key consequence of the exact, path-relative backward reachability test
\texttt{reach} is that the recursion never enters a region without outputs: every
non-initial branching call has an output within its subtree.

\begin{lemma}[No Fruitless Recursion]\label{lem:ss-no-fruitless}
  Every call other than the initial one yields at least one simple cycle through
  $s$ of length at most $k$ within its subtree of $T$.
\end{lemma}
\begin{proof}
  An output call yields such a cycle directly. Otherwise the call is a non-initial
  branching call, with current node $w=v_h\neq s$ (\autoref{obs:simple-path}); it is
  entered only because $w$ lay in the parent's \texttt{reached} set, i.e.\
  (\autoref{obs:reach}) $\mathrm{dist}_{V(P')}(w,s)\le \texttt{budget}'-1$ for the
  parent's path $P'$ and budget $\texttt{budget}'=k-\|P'\|$, with $w\notin V(P')$.
  Thus there is a simple $w$-$s$ path $W$ of length $\mathrm{dist}_{V(P')}(w,s)$
  whose interior avoids $V(P')$; prefixing $W$ with $P'$ yields a simple cycle $C$
  through $s$ of length
  $\|P'\|+1+\mathrm{dist}_{V(P')}(w,s)\le\|P'\|+1+(\texttt{budget}'-1)=k$, beginning
  with this call's prefix $(s,\dots,v_h)$. By \autoref{lem:ss-complete-call}, $C$ is
  yielded within this call's subtree of $T$.
\end{proof}

The initial call yields no output if and only if $G$ contains
no simple cycle through $s$ of length at most $k$.

Now, we bound the time between two consecutive outputs, as well as the time to the
first output and from the last output to termination. 

\begin{theorem}[Delay]\label{thm:ss-delay}
  For a simple directed graph $G$ with $n$ nodes and $m$ edges and length bound $0<k\le n$,
  $\textsc{simple\_search}(G,s,k)$ has per-output delay $O(k\,(n+m))$.
\end{theorem}
\begin{proof}
  When a branching call is \emph{entered} it issues one \texttt{reach}, costing
  $O(n+m)$ --- a breadth-first search visiting each node and edge at most once; its
  remaining work is $O(1)$ bookkeeping plus one pass over its \texttt{fruitful} list
  (length at most the out-degree of the current node), both dominated by that
  $O(n+m)$. An output call runs no \texttt{reach}; its cost beyond copying the
  emitted path is $O(1)$, and the copy is charged to the output. A \emph{return}
  from a call costs $O(1)$.

  List the output calls in execution order and bracket them by two virtual events at
  the root: let $o_0$ be the entry of the initial call, $o_{c+1}$ its return, and
  $o_1,\dots,o_c$ the $c$ output calls in between. By construction no output lies
  strictly between two consecutive events, and $o_0,o_{c+1}$ are the root. It
  suffices to bound the work in each interval $(o_j,o_{j+1})$, $0\le j\le c$: the
  interval $(o_0,o_1)$ is the time to the first output, the internal intervals are
  the inter-output delays, and $(o_c,o_{c+1})$ is the wrap-up from the last output to
  termination.

  Fix an interval $(o_j,o_{j+1})$. No call other than the initial one is entered and
  completed within it: by \autoref{lem:ss-no-fruitless} such a call would yield an
  output in its subtree, strictly between $o_j$ and $o_{j+1}$, contradicting that no
  output lies between these consecutive events. Hence the open-call stack turns at
  most once, at $a=\mathrm{lca}(o_j,o_{j+1})$: it ascends from $o_j$ to $a$ by
  returns alone, then descends from $a$ to $o_{j+1}$ by entries alone. (For $j=0$ the
  ascent is empty, as $a$ is the root; for $j=c$ the descent is empty.)

  The ascent is a sub-path of the root-to-$o_j$ path and the descent a sub-path of
  the root-to-$o_{j+1}$ path. The descent enters at most $k$ branching calls before
  reaching the output call $o_{j+1}$, contributing at most $k$ \texttt{reach} runs at
  $O(n+m)$ each; the ascent performs at most $k+1$ returns at $O(1)$ each. The work
  in the interval is thus $O(k\,(n+m))$, and the maximum over all intervals is the
  per-output delay $O(k\,(n+m))$.
\end{proof}

Up to the standard connectivity assumption $m\ge n-1$, under which $n+m=O(m)$,
this is the $O(k\,m)$ delay regime of Rizzi et al.~\cite{DBLP:journals/corr/RizziSS14}
for the related $st$-path listing problem.

\begin{corollary}[Total Time]\label{cor:ss-total}
  Let $c$ be the number of simple cycles through $s$ of length at most $k$ in $G$. 
  Then $\textsc{simple\_search}(G,s,k)$ runs in total time $O((c+1)\,k\,(n+m))$.
\end{corollary}
\begin{proof}
  There are $c$ output intervals and one final interval ending in termination,
  each of delay $O(k\,(n+m))$ by \autoref{thm:ss-delay}.
\end{proof}

\begin{theorem}[Space Complexity]\label{thm:ss-space}
  For a simple directed graph $G$ with $n$ nodes and $m$ edges, $\textsc{simple\_search}(G,s,k)$
  uses $O(n+m)$ working space, excluding the space charged to emitting outputs.
\end{theorem}
\begin{proof}
  The set computed by \texttt{reach} may be as large as the whole graph, but in
  the algorithm it is confined to the frame of \texttt{reach}: it is built there,
  consulted only to filter the current node's successors, and returned as that
  short \texttt{fruitful} list, so the set and its breadth-first frontier are gone
  before the enclosing call recurses. No reachable set is therefore live across a
  recursive descent.  What then persists from one search level to the next is, for each
  node on the current path, only its fruitful successors. Because the path is
  simple, its nodes are distinct, so the total length of these per-level lists
  cannot exceed the number of edges of the graph. Together with the single shared
  path and its membership set, and the scratch of the one breadth-first search
  active at any moment, the working space is linear in the size of the graph,
  independent of $k$ --- the space one expects of a depth-first traversal. This
  counts working space only; emitting a cycle copies the current path, which is
  charged to the output.
\end{proof}

\section{Conclusions and Outlook}

We have demonstrated that the algorithm \textsc{CYCLE\_SEARCH} proposed in~\cite{DBLP:journals/corr/abs-2105-10094} 
does not reliably enumerate all bounded-length simple cycles in a directed graph. 
We traced the omission to a single flaw in the algorithm's lock mechanism 
--- the reuse of a stack-relative backward distance under an incompatible search stack --- 
and pinpointed the corresponding invalid step in the original correctness proof. 
We have also demonstrated a gap in the original proof of the claimed delay bound,
which \textsc{SimpleSearch} avoids by admitting a direct per-output delay proof.
Finally, we presented a corrected algorithm, \textsc{SimpleSearch}, 
that avoids these flaws by construction and preserves the desirable property of 
having computational complexity sensitive to the specified cycle length.

In the course of our completeness testing on larger sets of small random graphs, 
we observed that the runtime per reported cycle of $\textsc{SimpleSearch}$ 
is in the same order of magnitude as that of \textsc{CYCLE\_SEARCH}.
Optimizing the implementation (including preprocessing and parallelization)
and rigorous experiments on specific graph sets are left to further studies.

In forthcoming work we study improved algorithms and alternative proof strategies for
both completeness and delay, established by methods different from those used here.

\section*{Acknowledgements}

The authors are grateful to Jochen Kerdels from HTW Berlin, Dept.\ of  Computer Engineering,
who independently implemented the algorithm in C++ and verified the counter-example.

\appendix
\section{Appendix: Python Code Demonstrating Incompleteness}\label{appendix:pseudocode}

The functions $\texttt{relax\_locks}$ and $\texttt{cycle\_search}$  
in \autoref{figure:pseudocode} below were adapted from~\cite{DBLP:journals/corr/abs-2105-10094} and implemented in Python.

\begin{figure}[H]
\small
\begin{lstlisting}[style=Python]
import math 
import networkx as nx

G = nx.parse_adjlist(['A D E', 'B D E', 'C A B', 'D A B', 'E C'], create_using=nx.DiGraph)

lock = {v : math.inf for v in G.nodes} 
Blist = {v: set() for v in G.nodes}
stack = []

def relax_locks(v, k, blen):
    if lock[v] < k-blen+1:
        lock[v] = k-blen+1
        for w in Blist[v]:
            if w not in stack:
                relax_locks(w, k, blen+1)
   
def cycle_search(G, v, k, flen):
    blen = math.inf
    lock[v] = flen
    stack.append(v)
    for w in G.successors(v):
        if w==stack[0]: # stack[0]==s
            print(f"new cycle of length {flen+1} found: {stack+[w]}")
            blen = 1
        elif (flen+1<lock[w]) and (flen+1<k):
            blen = min(blen, 1+cycle_search(G, w, k, flen+1))
    if blen < math.inf:
        relax_locks(v, k, blen)
    else:
        for w in G.successors(v):
            if v not in Blist[w]:
                Blist[w].add(v)
    stack.pop()
    return blen

cycle_search(G, 'A', 5, 0)              
\end{lstlisting}
\caption{Python Program demonstrating the Counter-Example}\label{figure:pseudocode}
\end{figure}

\section{Appendix: A Vertex Revisited Beyond the Stated Bound}\label{appendix:visit_bound}

The time complexity argument of~\cite{DBLP:journals/corr/abs-2105-10094} asserts that
no vertex is visited more than $k-1$ times before a valid cycle is detected. The
following instance violates this, and even the weaker bound of $k$ visits. We
instrumented $\texttt{cycle\_search}$ to count, for each vertex, the number of calls since
the most recent output (resetting all counts whenever a cycle is emitted). On the
strongly connected directed graph with edges
\[
\begin{array}{l}
(0,2),(0,3),(0,8),\;
(1,0),(1,2),\;
(2,0),(2,1),(2,5),\;
(3,4),(3,7),(3,8),\\
(4,6),(4,8),\;
(5,3),\;
(6,8),\;
(7,0),\;
(8,0),
\end{array}
\]
with $s=1$ and $k=8$, vertex $8$ violates the bound: between the outputs
$[1,0,2]$ and $[1,2]$ it is visited $9$ times --- exceeding both $k-1=7$ and
$k=8$. As the graph is strongly connected, the violation does not depend on any
artificial disconnection or sink structure. The instance was obtained by
searching small random graphs.

The mechanism is visible in the trace below; the leftmost column shows the search
stack, and only the output and the calls to $\texttt{cycle\_search}$ and $\texttt{relax\_locks}$ are
listed, in execution order. After the output $[1,0,2]$, the first deep search
exhausts a subtree; the subsequent $\texttt{relax\_locks}$ cascade walks back up a chain of
\emph{Blist} entries and raises the locks of several vertices --- vertex $8$ among
them, repeatedly --- which then re-admit it in sibling subtrees with no
intervening output. Its revisits accumulate accordingly until the next output
$[1,2]$.

\begin{figure}[H]
\small
\begin{lstlisting}
stack      function call
           cycle_search v=1 lock[v]=0
1          cycle_search v=0 lock[v]=1
10         cycle_search v=2 lock[v]=2
102        output [1, 0, 2]
102        cycle_search v=5 lock[v]=3
1025       cycle_search v=3 lock[v]=4
10253      cycle_search v=4 lock[v]=5
102534     cycle_search v=6 lock[v]=6
1025346    cycle_search v=8 lock[v]=7
102534     cycle_search v=8 lock[v]=6
10253      cycle_search v=7 lock[v]=5
10253      cycle_search v=8 lock[v]=5
102        relax_locks  v=2 lock[v]=8
10         cycle_search v=3 lock[v]=2
103        cycle_search v=4 lock[v]=3
1034       cycle_search v=6 lock[v]=4
1034       cycle_search v=8 lock[v]=4
103        cycle_search v=7 lock[v]=3
103        cycle_search v=8 lock[v]=3
10         cycle_search v=8 lock[v]=2
10         relax_locks  v=0 lock[v]=7
10         relax_locks  v=8 lock[v]=6
10         relax_locks  v=3 lock[v]=5
10         relax_locks  v=5 lock[v]=4
10         relax_locks  v=4 lock[v]=5
10         relax_locks  v=6 lock[v]=5
10         relax_locks  v=7 lock[v]=6
1          cycle_search v=2 lock[v]=1
12         cycle_search v=0 lock[v]=2
120        cycle_search v=3 lock[v]=3
1203       cycle_search v=4 lock[v]=4
12034      cycle_search v=8 lock[v]=5
1203       cycle_search v=7 lock[v]=4
1203       cycle_search v=8 lock[v]=4
120        cycle_search v=8 lock[v]=3
12         output [1, 2]
\end{lstlisting}
\caption{Execution trace on which vertex $8$ is visited by $\texttt{cycle\_search}$ nine times between the
outputs $[1,0,2]$ and $[1,2]$, exceeding the claimed bound of $k-1$ (here $k=8$).}\label{figure:visit_bound_trace}
\end{figure}

We do not pursue this further: the algorithm is already shown to be incomplete in
\autoref{section:counter_example}, and the observation here serves only to record
why we did not attempt to re-establish its delay bound.

\bibliographystyle{unsrtnat}
\bibliography{references}

@article{DBLP:journals/corr/RizziSS14,
  author       = {Romeo Rizzi and
                  Gustavo Sacomoto and
                  Marie{-}France Sagot},
  title        = {Efficiently listing bounded length st-paths},
  journal      = {CoRR},
  volume       = {abs/1411.6852},
  year         = {2014},
  url          = {http://arxiv.org/abs/1411.6852},
  eprinttype   = {arXiv},
  eprint       = {1411.6852},
  bibsource    = {dblp computer science bibliography, https://dblp.org}
}

@article{DBLP:journals/corr/abs-2105-10094,
  author       = {Anshul Gupta and
                  Toyotaro Suzumura},
  title        = {Finding All Bounded-Length Simple Cycles in a Directed Graph},
  journal      = {CoRR},
  volume       = {abs/2105.10094},
  year         = {2021},
  url          = {https://arxiv.org/abs/2105.10094},
  eprinttype    = {arXiv},
  eprint       = {2105.10094},
  bibsource    = {dblp computer science bibliography, https://dblp.org}
}

@article{DBLP:journals/siamcomp/Johnson75,
  author       = {Donald B. Johnson},
  title        = {Finding All the Elementary Circuits of a Directed Graph},
  journal      = {{SIAM} J. Comput.},
  volume       = {4},
  number       = {1},
  pages        = {77--84},
  year         = {1975},
  url          = {https://doi.org/10.1137/0204007},
  doi          = {10.1137/0204007},
  bibsource    = {dblp computer science bibliography, https://dblp.org}
}

@article{tiernan1970,
	author    = {J. C. Tiernan},
	title     = {An efficient search algorithm to find the elementary circuits of a graph},
	journal   = {Communications of the ACM},
	volume    = {13},
	number    = {12},
	pages     = {722--726},
  doi       = {10.1145/362814.362819},
  url       = {https://dl.acm.org/doi/10.1145/362814.362819},
	year      = {1970}
}

@article{tarjan1972,
  author = {Tarjan, Robert},
  title = {Depth-First Search and Linear Graph Algorithms},
  journal = {SIAM Journal on Computing},
  volume = {1},
  number = {2},
  pages = {146-160},
  year = {1972},
  doi = {10.1137/0201010},
  URL = {https://doi.org/10.1137/0201010},
  eprint = {https://doi.org/10.1137/0201010}
}

@article{szwarcfiter1976,
	author    = {J. L. Szwarcfiter and P. E. Lauer},
	title     = {A search strategy for the elementary cycles of a directed graph},
	journal   = {BIT Numerical Mathematics},
	volume    = {16},
	number    = {2},
	pages     = {192--204},
	year      = {1976},
  doi       = {10.1007/BF01931370},
  url       = {https://link.springer.com/article/10.1007/BF01931370}
}

@inproceedings{birmele2013,
	author    = {E. Birmelé and R. A. Ferreira and R. Grossi and A. Marino and N. Pisanti and R. Rizzi and G. Sacomoto},
	title     = {Optimal listing of cycles and st-paths in undirected graphs},
	booktitle = {Proceedings of the 24th Annual ACM-SIAM Symposium on Discrete Algorithms (SODA 2013)},
	pages     = {1884--1896},
  url       = {https://dl.acm.org/doi/10.5555/2627817.2627951},
  doi       = {10.5555/2627817.2627951},
	year      = {2013}
}

@inproceedings{DBLP:reference/algo/Grossi16,
  author="Grossi, Roberto",
  editor="Kao, Ming-Yang",
  title="Enumeration of Paths, Cycles, and Spanning Trees",
  bookTitle="Encyclopedia of Algorithms",
  year="2016",
  publisher="Springer New York",
  address="New York, NY",
  pages="640--645",
  isbn="978-1-4939-2864-4",
  doi="10.1007/978-1-4939-2864-4_728",
  url="https://doi.org/10.1007/978-1-4939-2864-4_728"
}
\end{document}